\documentclass[nocopyrightspace]{sigplanconf}

\usepackage{amsmath}
\usepackage{stmaryrd}
\usepackage{verbatim}
\usepackage{amsthm}
\usepackage{wasysym}
\usepackage{graphicx}
\usepackage{longtable}
\usepackage{hyperref}
\usepackage{breakurl}
\usepackage{color, colortbl}
\usepackage{paralist}

\newcommand{\nop}[1]{}
\def\punto{\hspace*{\fill}\Box}
\newtheorem{example}{Example}
\newtheorem{proposition}{Proposition}

\begin{document}



\title{Managing Schema Evolution in NoSQL Data Stores}

\authorinfo{Stefanie Scherzinger}
           {Regensburg University\break of Applied Sciences}
           {stefanie.scherzinger@hs-regensburg.de}
\authorinfo{Meike Klettke}
           {University of Rostock}
           {meike.klettke@uni-rostock.de}
\authorinfo{Uta St\"orl}
           { Darmstadt University\break of Applied Sciences}
           {uta.stoerl@h-da.de}

\maketitle

\begin{abstract}

NoSQL data stores are commonly schema-less,
providing no means for globally defining or managing the schema.
While this offers great flexibility in early stages of
application development, developers soon can experience the heavy
burden of dealing with increasingly heterogeneous data.
This paper targets schema evolution
for NoSQL data stores, the complex task of adapting and changing the
implicit structure of the data stored.
We discuss the re\-commendations of the developer community
on handling schema changes, and introduce a simple,
declarative schema evolution language.
With our language, software developers and architects
can systematically manage the evolution of their production data and
perform typical schema maintenance tasks.
We further provide a holistic NoSQL database programming language
to define the semantics of our schema evolution language.
Our solution does not require any modifications to the NoSQL data store,
treating the data store as a black box. Thus, we 
want to address application developers that use NoSQL systems
as database-as-a-service.

\end{abstract}






\section{Introduction}
\label{sec:introduction}

The classic database textbook dedicates several chapters to 
schema design: Carefully crafting an abstract model, 
translating it into a relational schema, which is then normalized.
While walking their students through the course, 
scholars emphasize again and again the importance of an anticipatory, 
holistic design, and the perils of making changes later on. Decades of 
experience in writing database applications have taught us this. 
Yet this waterfall approach no longer fits 
when building today's web applications.

During the last decade, we have seen radical changes in the way 
we build software, especially when it comes
to interactive, web-based applications: Release cycles have accelerated from yearly releases 
to weekly or even daily, new deployments of beacon 
applications such as Youtube (quoting Marissa Meyer in~\cite{lightstone}).
This goes hand in hand with developers striving to be agile.
In the spirit of lean development, design decisions are made as 
late as possible. This also applies to the schema. 
Fields that {\em might}\/ be needed  in the future are not added
presently,
reasoning that until the next release, things might change in a way 
that would render the fields unnecessary after all.
It is partly due to this very need for more flexibility, that 
schema-free NoSQL data stores
have become so popular. Typically, developers 
need not specify a schema up front. Moreover, adding a field
to a data structure can be done anytime and at ease.

\begin{figure}[h]
\centering

\includegraphics[scale=0.55]{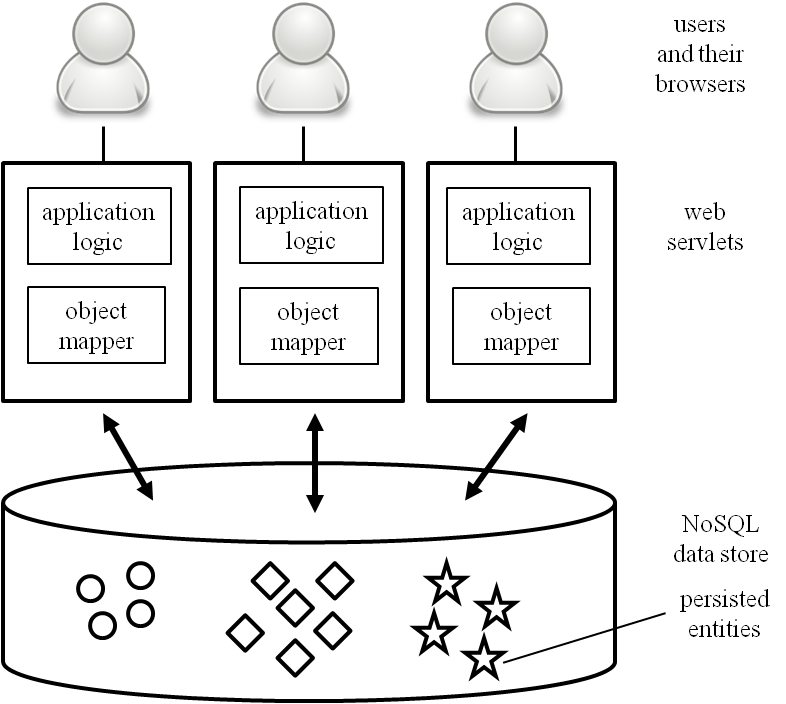}
\caption{Architecture of an interactive web application.}

\label{fig:architecture}

\end{figure}

\paragraph{Scope of this work.}
We study aspects of schema management for professional web applications
that are backed by NoSQL data stores. 
Figure~\ref{fig:architecture} sketches the typical architecture.
All users interact with
their own instance of the application, e.g.\ a servlet
hosted by  a platform-as-a-service, 
or any comparable web hosting service.
It is established engineering practice that the application code uses
an object mapper 
for the mapping of objects in the application space to 
the persisted entities.

We further assume that the NoSQL data store is provided
as database-as-a-service, so we have no way of configuring 
or extending it.
Our work addresses this important class of applications. Of course, there are
other use cases for employing NoSQL technology, 
yet they are not the focus of our work.

\paragraph{Case Study: Blogging applications.}

We  introduce a typical example of a professional
web application: An online blog. 
In the spirit of shipping early and often, 
both the features of our application as well as the data will evolve.

We use a NoSQL data store, which stores data as entities.
We will establish our terminology in later chapters,
and make do with a hand-wavy introduction at this point.
Each entity has an entity key, which is a tuple of an entity kind
and an identifier. Each entity further has a value,
which is a list of properties:

\begin{small}
\begin{verbatim}
(kind, id) = { comma-separated list of properties }
\end{verbatim}
\end{small}

Let us dive right in.
In our  first release, users publish blogs (with title and content) 
and guests can leave comments on blogposts. 
For each blogpost, information about the author and
 the date of the post is stored.
In the example below, we use a syntax inspired by JSON~\cite{json},
 a lightweight data-interchange format 
widely used within web-based applications.

\begin{small}
\begin{verbatim}
(blogpost, 007) = {
 title: "NoSQL Data Modeling Techniques",
 content: "NoSQL databases are often ...",
 author: "Michael",
 date: "2013-01-22",  
 comments: [ 
 { comment-content: "Thanks for the great article!",
   comment-date: "2013-01-24" },
 { comment-content: "I would like to mention ...",
   comment-date: "2013-01-26" } ] }
\end{verbatim}
\end{small}

Soon, we realize that changes are necessary: We decide to support voting, so that users
may ``like'' blogposts and comments. 
Consequently, we expand the structure of blogposts
and  add a ``likes'' counter.
Since we have observed some abuse, we  no longer support
 anonymous comments. From now on, users  authenticate 
 with a unique email address. Users may choose a username (\texttt{user}) and
link from their comments to their website (\texttt{url}), as well as
 specify  a list of interests. Accordingly, we add  new fields.

We  take a look at a new data store 
entity~\texttt{(blogpost, 234708)} and the state of an older
 entity \texttt{(blogpost, 007)} that has been added
in an earlier version of the application. 
For the sake of brevity,  we omit the data values:

\begin{small}
\begin{verbatim}
(blogpost, 234708) = {
 title, content, author, date, likes,
 comments [ 
  { comment-content, comment-date, comment-likes,
    user, email, url, interests[ ] } ] }
\end{verbatim}
\end{small}

\vspace{-0.3cm}
\begin{small}
\begin{verbatim}
(blogpost, 007) = {
 title, content, author, date, 
 comments [ 
  { comment-content, comment-date },
  { comment-content, comment-date } ] }
\end{verbatim}
\end{small}

Next, we decide to reorganize our user management.
We  store user-related data in separate \texttt{user} entities.
These entities contain the user's   \texttt{login}, \texttt{passwd},
 and \texttt{picture}. 

During this reorganization we further rename \texttt{email} to \texttt{login} 
in  \texttt{blogpost} entities. The  \texttt{interests} are moved from \texttt{blogpost}
to the \texttt{user} entities, and the \texttt{url} is removed.
Below, we show blogpost  \texttt{(blogpost, 331175)}
of this  new generation of data,
along with old generation blogposts that 
were persisted in earlier versions of the application.
The structural differences are apparent.

\begin{small}
\begin{verbatim}
(blogpost, 331175) = { 
 title, content, author, date, likes, 
 comments [
  { comment-content, comment-date, 
    comment-likes, user, login } ] }
(user, 42) = { login, passwd, interests[ ], picture }
\end{verbatim}
\end{small}

\vspace{-0.3cm}
\begin{small}
\begin{verbatim}
(blogpost, 234708) = {
 title, content, author, date, likes,
 comments [ 
  { comment-content, comment-date, comment-likes,
    user, email, url, interests[ ] } ] }
\end{verbatim}
\end{small}

\vspace{-0.3cm}
\begin{small}
\begin{verbatim}
(blogpost, 007) = {
 title, content, author, date, 
 comments [ 
  { comment-content, comment-date },
  { comment-content, comment-date } ] }
\end{verbatim}
\end{small}

After only three releases, we have accumulated considerable technical debt
in our application code.
It is now up to the developers to adapt their object mappers
and the application logic so that all three versions of blogposts may co-exist.

Whenever a blogpost is read from the data store,
the application logic has to account for the heterogeneity in the comments:
Some comments do not have any user information, while others have information 
about the user identified by \texttt{email} (along with other data).
A third kind of \texttt{blogpost} contains \texttt{comments} identified by the
user's \texttt{login}. 
%
If new generation comments are added to old generation blogposts, we produce even a fourth class of blogposts.

Not only does this introduce additional code complexity, it also increases
the testing effort.
%
 With additional case distinctions, a good code coverage
in testing becomes more difficult to obtain.

In an agile setting where
software is shipped early and often, developers would rather spend their
time writing new features than fighting such forms of technical debt.
At the same time, the NoSQL data store offers little, if any,
support in evolving the data along with the application.
Our main contribution in this paper is  an 
approach to solving these kinds of problems.


%

\paragraph{Schema evolution in schema-less stores.}
While the sweet spot of a schema-less backend is its 
flexibility, this freedom rapidly manifests in 
ever-increasing technical debt with growing
data structure entropy.
\nop{
Accessing a nonexistent property of an object may trigger
runtime exceptions. In the worst case, 
this exception propagates back to the user.
Ultimately, fear of such effects also increases the overall testing effort.
}
Once the data structures have degenerated, a NoSQL data store
provides little support for getting things straightened out.

Most NoSQL data stores do not offer a {\em data definition language}\/ 
for specifying a global schema (yet 
some systems, such as Cassandra, actually do~\cite{cassandra}).
Usually, they merely provide basic read- and write operations
for manipulating single entities,
delegating the manipulation of sets of entities
completely  to the application logic. 
Consequently, these systems offer no dedicated means
for migrating legacy entities, and developers
are referred to writing batch jobs for data migration tasks (e.g.~\cite{datastore_migration}). 
In such batch jobs, entities are fetched
one-by-one from the data store into the application
space, modified, and afterwards written back to the store.
Worse yet, since we consider interactive
web applications, migrations happen while the application is in use.
We refer to data migration in batches as {\em eager migration},
since entities are migrated in one go.

Alas, for a popular interactive web-application, the right moment for migrating
all entities  may never come.
Moreover, a large-scale data store may contain legacy data
that will never be accessed again, such as stale user accounts,
blogposts that have become outdated, or offers that have expired.
Migrating this data may be a wasted effort, and expensive,
when you are billed by your database-as-a-service provider
for all data store reads and writes.

As an alternative, the developer community pursues 
what we call a {\em lazy}\/ data migration strategy.
Entities of the old and new schema are allowed to co-exist.
Whenever an entity is read into the application space,
it can be migrated.
Effectively, this will migrate only ``hot'' data that is still 
relevant to users.
For instance, the Objectify object mapper~\cite{objectify_migration}
has such support for Google Datastore~\cite{google_datastore}. 
However, all structure manipulations require custom code.
As of today, there is no systematic way to statically analyze
manipulations before executing them. Moreover, from a database
theory point-of-view, 
 lazy migration is little understood (if at all).
This makes lazy migration a venture that, if applied
incorrectly on production data, poses great risks.
After all, once entities have been corrupted, there may be  no way to
undo the changes.

\paragraph{Desiderata.}
What is missing in today's  frameworks
is a means to systematically manage the schema of stored data,
while at the same time maintaining the flexibility that a schema-less
data store provides.
What we certainly cannot wish for is a rigorous corset
that ultimately enforces a relational schema on NoSQL data stores.

Most systems do provide some kind of data store viewer,
where single entities can be inspected, 
and even modified, or data can be deleted in bulk
(e.g.~\cite{datastore_viewer}).
Yet to the best of our knowledge, there is no schema management interface
that would work {\em across}\/ NoSQL systems from different providers,
allowing application administrators to manage their data's structure
systematically. This entails basic operations
such as adding or deleting fields, copying or moving fields
from one data structure to another.
From studying the discussions in developer forums, we have come to believe
that these are urgently needed operations
(e.g.\ \cite{datastore_migration,datastore_add_entity,objectify_migration} to list
just a few references).
Add, rename, and delete correspond to the capabilities
of an ``ALTER TABLE'' statement
in relational databases. Just as with relational databases,
more complex data migration tasks would then have to be encoded
programmatically.

Yet in the majority of NoSQL databases, {\em any}\/ data structure maintenance affecting more than one entity
must be coded manually~\cite{objectify_migration,datastore_migration, tiwari}.
We still lack some of the basic tooling that one would
expect in a NoSQL data store {\em ecosystem},
so that we may professionally maintain our production data in the long run.

\paragraph{Contributions.}
The goal of this work is to address this lack of tooling.
We lay the foundation for building a generic schema evolution  interface 
to NoSQL systems.
Such a tool is intended  for developers, administrators, and software architects
to declaratively manage the structure of their production data.
%
To this end,
we make the following contributions:

\begin{itemize}

\item We investigate the established field
of schema evolution  in the new context
of schema-less NoSQL data stores.

\item We contribute a declarative {\em NoSQL schema evolution language}\/.
Our language consists of a set of basic yet practical operations
that address the majority of the typical cases that we see discussed in developer forums.

\item 
We introduce a generic {\em NoSQL database programming language}\/
that abstracts from the APIs of the most prominent NoSQL systems.
Our language clearly distinguishes the state of the persisted data
from the state of the objects in the application space.
This is a vital aspect, since the NoSQL data store
offers a very restricted API, and data manipulation happens
 in the application code.

\item
By implementing our schema evolution operations
in our NoSQL database programming language, we show that
they can be implemented
for a large class of NoSQL data stores.

\item
We investigate whether a proposed  schema evolution operation 
is {\em safe}\/ to execute.

\item
Apart from exploring {\em eager}\/ migration,
we introduce the notion of {\em lazy}\/ migration
and point out its potential for future research
in the database community.

\end{itemize}

\paragraph{Structure.}
In the next section, we start with an
overview on the state-of-the-art in NoSQL data stores. 
Section~\ref{sec:evolution} introduces our declarative language for
evolving the data and its structure.
In Section~\ref{sec:api}, we define an abstract and generic 
NoSQL database programming language for accessing 
NoSQL data stores. The operations of our language are available in many
popular NoSQL systems. 
With this formal basis,
we can implement our schema evolution operations
eagerly, see Section~\ref{sec:encoding_evolution}.
Alternatively, schema evolution can be handled lazily.
We sketch the capabilities of object mappers that allow 
lazy migration in Section~\ref{sec:lazy}.
In Section~\ref{sec:related}, we discuss related work on schema evolution in 
relational databases, XML applications, and NoSQL data stores. 
We then conclude with a summary and an outlook on our future work.



\section{NoSQL Data Stores}
\label{sec:nosql}

We focus on NoSQL data stores hosted in a cloud environment.
Typically, such systems scale to  large amounts of data,
and are schema-less or schema-flexible.
We begin with a categorization of popular systems,
discussing their commonalities and differences.

We then point out the NoSQL data stores that
we consider in this paper with their core characteristics.
In doing so,  we generalize  from proprietary details
and introduce a common terminology.

\subsection{State of the art}

NoSQL data stores vary hugely in terms of data model, query model, 
scalability, architecture, and persistence design. 
Several taxonomies for NoSQL data stores
have been proposed.
Since we focus on schema evolution, a categorization of systems
by data  model is most natural for our purposes.
We thus resort to a (very common) classification~\cite{cattell, tiwari}
into (1)~key-value stores, 
(2)~document stores, and (3)~extensible record stores.
Often, extensible record stores are also called 
wide column stores or column family stores.

\paragraph{(1) Key-value stores.}
Systems like Redis~\cite[Chapter~8]{seven} or Riak~\cite{riak} store data in pairs 
of a unique key and a value. 
Key-value stores do not manage the structure of these values.
There is no concept of schema beyond distinguishing keys and values. 
Accordingly, the query model is very basic:
Only inserts, updates, and deletes by key are supported,
yet no query predicates on values. 
Since key-value stores do not manage the schema of values,
schema evolution is the responsibility of 
the application.

\paragraph{(2) Document stores.}
Systems such as MongoDB~\cite{mongodb} or Couchbase~\cite{couchbase}
also store key-value pairs. However, they store ``documents'' in the 
value part. The term ``document'' 
connotes loosely structured sets of name-value pairs, typically 
in JSON (JavaScript Object Notation) format or the binary representation
BSON, a more type-rich format of JSON.
Name-value pairs represent the properties of data objects. 
Names are unique, and name-value pairs are sometimes even referred to as 
key-value pairs.
The document format is hierarchical, so values may be
scalar, lists, or even nested documents. 
Documents within the same document store may differ 
in their structure, since there is no fixed schema. 

Queries in document stores are more expressive than in key-value stores. 
Apart from inserting, updating, and deleting documents based on the 
document key, we may query documents based on their properties.
The query languages differ from system to system.
Some systems, such as MongoDB, 
have an integrated query language for
ad-hoc queries, whereas other systems, such as CouchDB~\cite[Chapter~6]{seven} 
and Couchbase, do not. 
There, the  user predefines views
in form of MapReduce functions~\cite{mapreduce, tiwari}.

An interesting and orthogonal point
is the behavior in evaluating predicate queries: 
When a document does not contain a property mentioned in a query
predicate, then this property is not even considered in query evaluation.

Document stores are schema-less, so documents may effortlessly evolve in structure:
Properties can be added or removed from a particular document
without affecting the remaining documents. 
Typically, there is no schema definition language
that would allow the application developer to manage the 
structure of documents globally, across all documents.

\paragraph{(3) Extensible record stores.}
Extensible record stores such as  BigTable~\cite{bigtable}
or HBase~\cite{hbase} actually provide a loosely defined schema.
Data is stored as records. A schema defines
families of properties, and new properties can be added within 
a property family on a per-record basis. (Properties and property families are often also referred to
 as {\em columns}\/ and {\em column families}.)
Typically, the schema cannot be defined up front and extensible record stores 
allow the ad-hoc creation of new properties. 
However, properties cannot be renamed or easily re-assigned from 
one property family to the other. 
So certain challenges from schema evolution in relational database systems
carry over to extensible record stores.

Google Datastore~\cite{google_datastore} is
built on top of Megastore~\cite{megastore} and BigTable,
and is very flexible and comfortable to use.
For instance, it very effectively implements multitenancy
for all its users.

The Cassandra system~\cite{cassandra} is an exception
among extensible record stores,
since it is much more restrictive regarding schema. 
Properties are actually defined up front, even with a ``CREATE TABLE'' statement,
and the schema is altered globally with an ``ALTER TABLE'' 
statement.
So while Cassandra is an extensible record store~\cite{cattell, tiwari},
it is not schema-less or schema-flexible.
In this work, we will exclusively consider  schema-less data stores.

\paragraph{A word on NULL values.}
The handling of NULL values in NoSQL data stores deserves attention,
as the treatment of unknown values is a factor in schema evolution.
In relational database systems, NULL values represent unknown information,
and are processed with a three-valued logic in query evaluation.
Yet in NoSQL data stores, there is no common notion of NULLs across systems:
\begin{itemize}

\item Some systems follow the same semantics of NULL values as relational databases,
e.g.~\cite{mongodb}.

\item Some systems allow for NULL values to be stored,
but do not allow NULLs in query predicates,
e.g.~\cite{cassandra,google_datastore}.

\item Some systems do not allow NULL values at all, e.g.~\cite{hbase},
arguing that NULL values only waste storage.

\end{itemize}
While there is no common strategy on handling unknown values yet,
the discussion is ongoing and lively. Obviously, 
there is a semantic difference between a property value that is not known
(such as the first name for a particular user),
and a property value that does not exist for a variant of an entity
(since home addresses and business addresses are structured differently).
Consequently, some NoSQL data stores which formerly did not support NULL values
have introduced them in later releases~\cite[Chapter~6]{seven, couchpotato}.

In Section~\ref{sec:api}, we present a generic NoSQL data store
programming language. As the approaches to handling NULL values
are so manifold, we choose to disregard NULLs as values and in queries,
until a consensus has been established  among NoSQL data stores.

\subsection{NoSQL Data Stores in Scope for this Paper}

In this paper, we investigate schema evolution
for feature-rich, interactive web applications
that are backed by NoSQL data stores.
This makes document stores and schema-less extensible record stores 
our primary platforms of interest.
Since key-value stores do not know any schema apart from distinguishing keys and values,
we believe they are not the  technology of choice for 
our purposes; after all, one cannot even run the most basic predicate queries,
e.g.\ to find all blogs posted within the last ten hours.

We assume a high-level, abstract view on document stores and extensible record stores
and  introduce our terminology.
Our terminology takes after Google Datastore~\cite{google_datastore}.
We also state  our assumptions on the data and query model.

\label{sec:terminology}
\paragraph{Data model.} 
Objects  stored in the NoSQL data store are called {\em entities}\/.
Each entity belongs to a {\em kind}\/, which is a name given to 
groups of semantically similar objects.
Queries can then be specified over all entities of the same kind.
Each entity  has a unique {\em key}\/, which consists of the entity kind 
and an {\em id}\/. 
Entities have several {\em properties}\/ (corresponding to attributes
in the relational world).
Each entity property consists of a {\em name}\/ and a {\em value}.
Properties may be scalar, they may be multi-valued, or consist of nested entities.

\paragraph{Query model.} 
Entities can be inserted and deleted based on their key.
We can formulate queries against all entities of a kind.
At the very least, we assume that a NoSQL data store
supports conjunctive queries with equality comparisons.
This functionality is commonly provided by document stores 
and extensible record stores alike.

\paragraph{Freedom of schema.} 
We assume that the global structure of entities cannot be fixed in advance.
The structure of a single entity can be changed any time,
 according to the developers' needs.

\begin{example} \em
The blogging application example from  the Introduction 
is coherent with this
terminology and these assumptions.
$\punto$
\end{example}



\section{A NoSQL Schema Evolution Language}
\label{sec:evolution}

In schema-less NoSQL data stores, 
there is no explicit, global schema.
Yet when we are building feature-rich, interactive web applications on top of 
NoSQL data stores, 
entities actually do display  an implicit structure (or schema); 
this structure manifests in  the entity kind
and entity property names.
This especially holds when  object mappers 
take over the mundane task
of marshalling objects from the application space
into persisted entities, and back. These object mappers commonly map class names
to entity kinds, and class members to entity properties.
(We discuss object mappers further in the context of related work
in Section~\ref{sec:related}.)

Thus, there is a large class of applications that use NoSQL data stores,
where the data is {\em somewhat}\/ consistently structured, but has no fixed
schema in the relational sense. Moreover, in an agile setting,
these are applications that evolve rapidly, both in their features
and their data.
Under these assumptions, we now define a compact set of declarative
schema migration operations, that have been inspired by
schema evolution in  relational databases,
and update operations for semi-structured 
data~\cite{xquery-update}. While we can only argue empirically,
having read through discussions in various developer forums,
we are confident that these operations cover a large share of the common 
schema migration tasks.

\begin{figure}[tb]

\noindent

\small
\begin{verbatim}
evolutionop ::= add | delete | rename | move | copy;
 
add ::= "add" property "=" value [selection]; 
delete ::= "delete" property [selection]; 
rename ::= "rename" property "to" pname [selection]; 
move ::= "move" property "to" kname [complexcond];
copy ::= "copy" property "to" kname [complexcond];

selection ::= "where" conds;
complexcond ::= "where"  (joincond | conds  
                          | (joincond "and" conds));
joincond ::= property "=" property;
conds ::= cond {"and" cond};
cond ::=  property "=" value;

property ::= kname "." pname;
kname ::= identifier;
pname ::= identifier;
\end{verbatim}

\nop{propertyspec ::= kindname  "." propertyname 
                 [condition] { ( "." propertyname ) } ;

condition ::= "[" propertyname "=" value "]";
}

\caption{EBNF of the NoSQL schema evolution language.}

\label{fig:ebnf}
\end{figure}

Figure \ref{fig:ebnf} shows 
the syntax of our {\em NoSQL schema evolution language}\/ in Extended Backus-Naur Form (EBNF).
An evolution operation adds, deletes, or renames properties.
Properties can also be moved or copied. Operations may contain
conditionals, even joins. The property kinds (derived from \verb!kname!)
and the property names (\verb!pname!) are the terminals in this grammar.
We will formally specify the semantics for our operations
in Section~\ref{sec:encoding_evolution}. For now, we discuss some examples
to develop an intuition for this language. 

We introduce a special-purpose numeric property~``version'' for all entities.
The version is incremented each time an entity is processed
by an evolution operator. This allows us to manage
heterogeneous entities of the same kind. This is an established development practice
in entity evolution.

\noindent
We begin with operations that affect all entities of one kind:
\begin{compactitem}

\item The add operation adds a property
to all entities of a given kind.
A default value may be specified  (see Example~\ref{ex:add}).

\item The delete operation removes a property 
from all entities of a given kind (see Example~\ref{ex:delete}).

\item The rename operation changes the name of a property 
for all entities of a given kind (see Example~\ref{ex:rename}).

\end{compactitem}

\begin{example} \em \label{ex:add}
Below, we show an entity from our
blogpost example before and after applying operation
{\bf add blogpost.likes = 0}. This adds
a likes-counter to all blogposts, initialized to zero.
We chose a compact tabular representation of entities and their properties.

\noindent
{\flushleft
\setlength{\tabcolsep}{.35em}
\begin{tabular}{c  c|c c}
\footnotesize
\begin{tabular}[t]{|l|l|}
\hline
{\bf key} & {\bf (blogpost, 331175)} \\
\hline
title & NoSQL Data.. \\
\hline
content & NoSQL databases  ..\\
\hline
{\em version} & {\em 1} \\
\hline
\end{tabular}
&
\quad
&
\quad
&
\footnotesize
\begin{tabular}[t]{|l|l|}
\hline
{\bf key} & {\bf (blogpost, 331175)} \\
\hline
title & NoSQL Data.. \\
\hline
content & NoSQL databases..\\
\hline
likes & 0 \\
\hline
{\em version} & {\em 2} \\
\hline
\end{tabular}
\end{tabular}
}

\vspace{-1.9mm}
$\punto$
\end{example}

\begin{example} \em \label{ex:delete}
The operation {\bf {delete blogpost.url}}
deletes the property ``url'' from all blogposts.
\noindent
{\flushleft
\setlength{\tabcolsep}{.35em}
\begin{tabular}{c  c|c c}
\footnotesize
\begin{tabular}[t]{|l|l|}
\hline
{\bf key} & {\bf (blogpost, 331175)} \\
\hline
title & NoSQL Data.. \\
\hline
content & NoSQL databases ..\\
\hline
url & www.mypage.com\\
\hline
{\em version} & {\em 1} \\
\hline
\end{tabular}
&
\quad
& 
\quad
&
\footnotesize
\begin{tabular}[t]{|l|l|}
\hline
{\bf key} & {\bf (blogpost, 331175)} \\
\hline
title & NoSQL Data.. \\
\hline
content & NoSQL databases..\\
\hline
{\em version} & {\em 2} \\
\hline
\end{tabular}
\end{tabular}
}\\

\noindent
We can also specify a selection predicate.
For instance,  {\bf delete blogpost.url where blogpost.version = 1}
deletes the url-property only from those blogposts
that are at schema version~1.
$\punto$
\end{example}

\begin{example} \em \label{ex:rename}
{\bf {rename blogpost.text to content}} renames
 the property ``text'' to ``content''
for all blogpost entities.

\noindent
{\flushleft
\setlength{\tabcolsep}{.35em}
\begin{tabular}{c  c|c c}
\footnotesize
\begin{tabular}[t]{|l|l|}
\hline
{\bf key} & {\bf (blogpost, 331175)} \\
\hline
title & NoSQL Data.. \\
\hline
text & NoSQL databases..\\
\hline
{\em version} & {\em 1} \\
\hline
\end{tabular}
&
\quad
&
\quad
&
\footnotesize
\noindent\begin{tabular}[t]{|l|l|}
\hline
{\bf key} & {\bf (blogpost, 331175)} \\
\hline
title & NoSQL Data.. \\
\hline
content & NoSQL databases..\\
\hline
{\em version} & {\em 2} \\
\hline
\end{tabular}
\end{tabular}
}

\vspace{-1.9mm}
$\punto$
\end{example}

We define further operations that affect two kinds of entities.
Such migration operations are not available in schema definition languages
for relational databases. Yet since NoSQL data stores typically do not support
joins, denormalization is a technique heavily relied upon. 
When building interactive
web applications, responsiveness is key,
which usually forbids programmatic joins in the application.
Instead,
one would reorganize that data such that it renders joins unnecessary.
Thus,  duplication and denormalization are first-class citizens when
building applications on top of NoSQL data stores. 
Accordingly, we introduce
dedicated operations for supporting these schema refactorings.

\begin{compactitem}

\item The move operation moves a property from one entity-kind
to another entity-kind (see Example~\ref{ex:move}).

\item The copy operation copies a property 
from one entity-kind to another entity-kind (see Example~\ref{ex:copy}).

\end{compactitem}
Of course, in moving and copying we also compute joins. Yet  this is done
in offline batch processing, and not during time-critical
interactions with users.

\nop{
\paragraph{Move:} In contrast to relational databases, we additionally want to support a \texttt{move} and \texttt{copy}
operation for evolving structures. In most NoSQL data stores, the queries not support join operations. For that reason, 
users either have to realize data join manually by programming nested loops in the application  or store (for reasons of efficiency) data in a redundant form. 
NoSQL databases often do not normalize data, we can observe a paradigm shift with respect to the design of applications.
Duplication and denormalization are first-class citizens in NoSQL systems. 
}

\begin{example} \em \label{ex:move}
To move the property ``url'' from users to all their blogposts,
we specify the operation  {\bf move user.url to blogpost
where user.name = blogpost.author}.
Figure~\ref{fig:move} shows its application to a  blog by user Gerhard.
$\punto$
\end{example}

\begin{figure}

\noindent
{\flushleft
\setlength{\tabcolsep}{.35em}
\begin{tabular}{l l|l l}

\footnotesize
\begin{tabular}[t]{|l|l|}
\hline
{\bf key} & {\bf (user, 1234)}\\
\hline
name & Gerhard \\
\hline
 email & gerhard@acm.org\\
\hline
status & professional\\
\hline
url & http://bigdata.org\\
\hline
{\em version} & {\em 1} \\
\hline
\end{tabular}

& & &

\footnotesize
\begin{tabular}[t]{|l|l|}
\hline
{\bf key} & {\bf (user, 1234)}\\
\hline
name & Gerhard \\
\hline
 email & gerhard@acm.org\\
\hline
status & professional\\
\hline
{\em version} & {\em 2} \\
\hline
\end{tabular}

\\

\footnotesize
\begin{tabular}[t]{|l|l|}
\hline
{\bf key} & {\bf (blogpost, 331175)} \\
\hline
title & NoSQL Data .. \\
\hline
content & NoSQL databases..\\
\hline
author & Gerhard \\
\hline
{\em version} & {\em 1} \\
\hline
\end{tabular}

&
\quad
&
\quad
&

\footnotesize
\begin{tabular}[t]{|l|l|}
\hline
{\bf key} & {\bf (blogpost, 331175)} \\
\hline
title & NoSQL Data.. \\
\hline
content & NoSQL databases..\\
\hline
author & Gerhard \\
\hline
url & http://bigdata.org\\
\hline
{\em version} & {\em 2} \\
\hline
\end{tabular}

\end{tabular}

}

\caption{Moving property ``url'' (c.f.\ Example~\ref{ex:move}).}
\label{fig:move}
\end{figure}

\begin{example} \em \label{ex:copy}
The next example shows the copy operation:
The property ``email'' is copied from users to all their blogposts: 
{\bf copy user.email to blogpost where user.name = blogpost.author}.
Figure~\ref{fig:copy} shows its application to a blog by user Gerhard. The copy operation does not change 
the user entities.
$\punto$
\end{example}

\begin{figure}
{\flushleft
\setlength{\tabcolsep}{.35em}
\begin{tabular}{l l|l l}

\footnotesize
\begin{tabular}[t]{|l|l|}
\hline
{\bf key} & {\bf (user, 1234)}\\
\hline
name & Gerhard \\
\hline
 email & gerhard@acm.org\\
\hline
status & professional\\
\hline
{\em version} & {\em 1} \\
\hline
\end{tabular}

& \quad & \quad &

\footnotesize
\begin{tabular}[t]{|l|l|}
\hline
{\bf key} & {\bf (user, 1234)}\\
\hline
name & Gerhard \\
\hline
 email & gerhard@acm.org  \\
\hline
status & professional \\
\hline
{\em version} & {\em 1} \\
\hline
\end{tabular}

\\

\footnotesize
\begin{tabular}[t]{|l|l|}
\hline
{\bf key} & {\bf (blogpost, 331175)} \\
\hline
title & NoSQL Data  .. \\
\hline
content & NoSQL databases  ..\\
\hline
author& Gerhard\\
\hline
{\em version} & {\em 1} \\
\hline
\end{tabular}

& \quad & \quad &

\footnotesize
\begin{tabular}[t]{|l|l|}
\hline
{\bf key} & {\bf (blogpost, 331175)} \\
\hline
title & NoSQL Data  .. \\
\hline
content & NoSQL databases  ..\\
\hline
author& Gerhard\\
\hline
 email & gerhard@acm.org\\
\hline
{\em version} & {\em 2} \\
\hline
\end{tabular}

\end{tabular}

}

\caption{Copying property ``email'' (c.f.\ Example~\ref{ex:copy}).}
\label{fig:copy}
\end{figure}

Section \ref{sec:encoding_evolution} formalizes the semantics and
investigates the  effort of our migration operations.
As a prerequisite, we next introduce a generic NoSQL database programming language.



\section{A NoSQL Database Programming Language}
\label{sec:api}

Relational databases come with a query language capable of joins,
as well as dedicated data definition and data manipulation language.
Yet in programming against NoSQL data stores, the application logic
needs to take over some of these responsibilities.
We now define the typical operations on entities in NoSQL data stores,
building a purposeful NoSQL database programming language.
Our language is particularly modeled after the interfaces to 
Google Datastore~\cite{google_datastore}, and is applicable
to document stores (e.g.~\cite{couchbase}) as well as schema-less extensible 
record stores (e.g.~\cite{hbase}). 

We consider system architectures such as shown in Figure~\ref{fig:architecture}.
Each user interacts
with an instance of the application, e.g.\ a servlet.
Typically, the application fetches entities from the data store into the application space,
modifies them, and writes them back to the data store.
We introduce a common abstraction from the current state 
of the data store and the objects available in the application space.
We refer to this abstraction as the {\em memory state}\/.

\paragraph{The memory state.}
We model a memory state as a set of mappings
from entity keys to entity values. 
Let us assume that an entity
has key~$\kappa$ and value~$\vartheta$.
Then the memory contains the mapping from this key to this value: $\kappa \mapsto \vartheta$.
Keys in a mapping are unique, so a memory state does not contain any mappings
$\kappa \mapsto \vartheta_1$ and~$\kappa \mapsto \vartheta_2$
with~$\vartheta_1 \neq \vartheta_2$.

The entity value itself is structured
as a mapping from property names to property values. 
A property value may be from an atomic domain~$\textit{Dom}$,
either single-valued ($\textit{Dom}$) or multi-valued ($\textit{Dom}^+$),
or it may consist of the properties of a nested entity.

\begin{example} \em \label{ex:single_entity}
We model a memory state with a single entity managing user data.
The key is a tuple  of kind
{\em user}\/ and the id~$42$. 
The entity value contains the user's login~``hhiker'' and password~``galaxy'':
$\{ (\textit{``user''}, 42) \mapsto \{\textit{login} \mapsto \textit{``hhiker''}, \;
\textit{pwd} \mapsto \textit{``galaxy''}\}  \}$.
$\punto$
\end{example}

\paragraph{Substitutions.} We describe manipulations of a memory state
by substitutions. A substitution~$\sigma$ is a mapping from a set~$K$ (e.g.\ the entity keys)
to a set~$V$ (e.g.\ the entity values) and the special symbol~$\bot$.
To access~$\vartheta_i$ in a substitution
$\{\kappa_1 \mapsto \vartheta_1, \dots, \kappa_n \mapsto \vartheta_n\}$, 
we write~$\sigma(\kappa_i)$.
If $\sigma(\kappa_i) = \bot$, then this explicitly means that 
this mapping is not defined.
%

Let~$\textit{ms}$ be the memory state, and let~$\sigma$ be a substitution.
In {\em updating the memory state $\textit{ms}$ by substitution~$\sigma$}\/,
we follow a create-or-replace philosophy for each mapping 
in the substitution.
We denote the updated memory by $\textit{ms}[\sigma]$:
$$\textit{ms}[\sigma] = \bigcup_{m \in \textit{ms}}(m[\sigma]).$$
\noindent
Let $a, b \in K$, and let $v, w \in V \cup \{\bot\}$. Then
\begin{align*}
&\{a \mapsto v\} [\sigma]
= 
\bigcup_{\{b \mapsto w\} \in \sigma} (\{a \mapsto v\}[\{b \mapsto w\}])
\\
&\{a \mapsto v\}[\{b \mapsto w\}] =
\left\{ 
  \begin{array}{l l}
   \{b \mapsto w\} & a = b\\
   \{a \mapsto v, b \mapsto w\}  &  \text{otherwise}
  \end{array} \right.
\end{align*}

We further  use the shortform 
$\textit{ms}[\kappa \mapsto \vartheta]$ to abbreviate the 
substitution with a single mapping~$\textit{ms}[\{\kappa \mapsto \vartheta\}]$.

\begin{example} \em
We continue with Example~\ref{ex:single_entity} and abbreviate the 
key~$(\textit{``user''}, 42)$ by~$k$. To 
delete the user's account, we update the memory state as 
\begin{align*}
&\{ k \mapsto \{\textit{login} \mapsto \textit{``hhiker''}, 
\textit{pwd} \mapsto \textit{``galaxy''}\}  \}  [k \mapsto \bot] \\
& =  \{k \mapsto \bot\}.
\end{align*}

\noindent
To mark the account as expired (state ``x''), we write
\begin{align*}
&\{ k \mapsto (\{\textit{login} \mapsto \textit{``hhiker''}, \;
\textit{pwd} \mapsto \textit{``galaxy''}\} [\textit{state} \mapsto \textit{``x''}]) \} 
\\
&= \{ k \mapsto \{\textit{login} \mapsto \textit{``hhiker''}, \;
\textit{pwd} \mapsto \textit{``galaxy''}, \textit{state} \mapsto \textit{``x''}\}\}.
\end{align*}
To change the user's password to ``g2g'', we write
\begin{align*}
&\{ k \mapsto (\{\textit{login} \mapsto \textit{``hhiker''}, \;
\textit{pwd} \mapsto \textit{``galaxy''}\} [\textit{pwd} \mapsto \textit{``g2g''}] ) \} 
\\
&= \{ k \mapsto \{\textit{login} \mapsto \textit{``hhiker''}, \;
\textit{pwd} \mapsto \textit{``g2g''}\}\}.
\end{align*}
\vspace{-5mm}
$\punto$
\end{example}

\paragraph{Evaluating operations.}
Operations may change the state of the data store and the application space.
We call the former the {\em data store state}\/,
 and call the latter the {\em application state}\/.
We denote the impact of operations by rules of the form
$$
\llbracket \textit{op} \rrbracket (\textit{ds}, \textit{as}) 
= (\textit{ds}', \textit{as}')
$$
where $\textit{op}$ denotes the operation to be executed
on the data store state~$\textit{ds}$ and the application state~$\textit{as}$.
By evaluating the operation, the data store state changes 
to $\textit{ds}'$,
and the application state to $\textit{as}'$.

\noindent
Operations may be executed in sequence,
which we define as
$$
\llbracket \textit{op}_1; \;\textit{op}_2 \rrbracket(\textit{ds}, \textit{as}) =
\llbracket  \textit{op}_2 \rrbracket \big( \llbracket \textit{op}_1 \rrbracket 
(\textit{ds}, \textit{as})  \big).
$$

\begin{figure*}
Let $\textit{ds}$ and $\textit{as}$ be a data store state and an application state.
Let $\kappa, \kappa'$ be entity keys.   
Let $n, n'$ be property names, and 
let $v$ be a property value.
The symbol~$\bot$ denotes the undefined value.
Let $\pi, \pi'$ be properties, i.e.\ a set of mappings from property names to property values.
$\textit{kind}: \textit{Keys} \mapsto \textit{Kind}$
is a function that extracts the entity kind from a key. $\theta$ is a conjunctive query,
and $c$ is a string constant.
\begin{align}
\llbracket \mbox{new}(\kappa) \rrbracket(\textit{ds}, \textit{as}) 
& =  (\textit{ds}, \textit{as}[\kappa \mapsto \emptyset])
\label{sem:new}\\
\llbracket \mbox{new}(\kappa, \pi) \rrbracket(\textit{ds}, \textit{as}) 
& =  (\textit{ds}, \textit{as}[\kappa \mapsto \pi])
\label{sem:new_initialized}\\
 \llbracket \mbox{setProperty}(\kappa, n, v) \rrbracket(\textit{ds}, \textit{as} \cup \{\kappa \mapsto \pi\}) 
& = (\textit{ds}, \textit{as} \cup \{\kappa \mapsto (\pi[n \mapsto v])\})
\label{sem:setproperty} \\
 \llbracket \mbox{setProperty}(\kappa, n, \kappa') \rrbracket(\textit{ds}, \textit{as} \cup \{\kappa \mapsto \pi\}
\cup \{\kappa' \mapsto \pi'\}) 
& = (\textit{ds}, \textit{as} \cup \{\kappa \mapsto (\pi[n \mapsto \pi'])\} \cup \{\kappa' \mapsto \pi'\})
\label{sem:setproperty_nested} \\
 \llbracket \mbox{removeProperty}(\kappa, n) \rrbracket(\textit{ds}, \textit{as} \cup \{\kappa \mapsto \pi\}) 
& = (\textit{ds}, \textit{as} \cup \{\kappa \mapsto (\pi[n \mapsto \bot])\})
\label{sem:removeproperty} \\[1mm]
\llbracket \mbox{put}(\kappa) \rrbracket(\textit{ds}, \textit{as} \cup \{\kappa \mapsto \pi\}) 
& = (\textit{ds} [\kappa \mapsto \pi], \textit{as} \cup \{\kappa \mapsto \pi\})
\label{sem:put} \\
\llbracket \mbox{delete}(\kappa) \rrbracket(\textit{ds}, \textit{as}) 
& = (\textit{ds} [\kappa \mapsto \bot], \textit{as})
\label{sem:delete} \\
\llbracket \mbox{get}(\kappa) \rrbracket(\textit{ds} \cup \{\kappa \mapsto \pi\}, \textit{as}) 
& = (\textit{ds} \cup \{\kappa \mapsto \pi\}, \textit{as} [\kappa \mapsto \pi])
\label{sem:get} \\[1mm]
%
\llbracket \mbox{get}(\textit{kind} = c) \rrbracket(\textit{ds}, \textit{as})
&  =  (\textit{ds}, \textit{as} [ \{\kappa \mapsto \pi \mid \kappa \mapsto \pi \in \textit{ds} \land
\textit{kind}(\kappa) = c\}])
\label{sem:query_kind}\\
\llbracket \mbox{get}(\textit{kind} = c \land \theta) \rrbracket(\textit{ds}, \textit{as}) 
& = (\textit{ds}, as [ \{ \kappa \mapsto \pi \mid 
 \kappa \mapsto \pi \in \textit{ds} \land
\textit{kind}(\kappa) = c \land
\llbracket \theta \rrbracket(\kappa \mapsto \pi)
\}])
\label{sem:query_theta} 
\end{align}

\caption{Operations for 
creating, manipulating, and persisting entities, as well as queries over entities.}
\label{fig:semantics}
\end{figure*}

\subsection{Manipulating Entities}
\label{sec:manipulating_entities}

We  next formalize operations common to most NoSQL data stores, namely 
creating and persisting entities, as well as retrieving and deleting single entities.
Figure~\ref{fig:semantics} defines our operations.
Let $\textit{Kind}$ be the set of entity kinds. Let $\textit{Id}$ be a set denoting identifiers.
The set of entity keys is defined as $\textit{Keys} = \textit{Kind} \times \textit{Id}$,
i.e.\ an entity key is a tuple of the kind and an identifier.
Entity properties are named. Let $\textit{Names}$ be the set of property names.
A property value can be either an atomic value from domain $\textit{Dom}$,
multi-valued (i.e.\ from $\textit{Dom}^+$), or a nested entity.

\paragraph{Creating entities.}
We start with the rules to create entities and their properties.
They affect the application state only. (For a change to have lasting effect,
the entity must be persisted.)
Rule~\ref{sem:new} creates a new entity with key~$\kappa$.
Initially, an entity does not have any properties. To also set initial properties,
we can use rule~\ref{sem:new_initialized}.
Rule~\ref{sem:setproperty} adds a new property with name~$n$ and value~$v$
to the entity with key~$\kappa$.
Adding a nested entity as a property is specified in Rule~\ref{sem:setproperty_nested}.
Rule~\ref{sem:removeproperty} removes the property with name~$n$
from the entity with key~$\kappa$: By setting the property value to~$\bot$,
the property by that name is no longer defined.

\paragraph{Persisting entities.}
Rule~\ref{sem:put} persists the entity with key~$\kappa$, 
replicating this entity to the data store state.
The put-operation replaces any entity by the same key,
should one exist. Rule~\ref{sem:delete} deletes the entity with key~$\kappa$ 
from the data store state.
With rule~\ref{sem:get}, we retrieve a particular entity by key
from the data store state.

\begin{example}\em 
The following sequence of operations creates the entity from Example~\ref{ex:single_entity}
and persists it in the data store.
\begin{tabbing}
xxx \= xxxxxx \= \kill
\> \textcircled{\raisebox{-0.9pt}{1}}  \>    new\big((``{\em user}'', $42$)\big); \\
\> \textcircled{\raisebox{-0.9pt}{2}}  \> setProperty\big((``{\em user}'', $42$),  {\em login}, ``{\em hhiker}''\big); \\
\> \textcircled{\raisebox{-0.9pt}{3}} \> setProperty\big((``{\em user}'', $42$),  {\em pwd},  ``{\em galaxy}''\big); \\
\> \textcircled{\raisebox{-0.9pt}{4}} \> put\big((``{\em user}'', $42$)\big)
\end{tabbing}
\noindent
We evaluate the operations one by one on the initially
empty data store and application state:
\begin{align*}
 & (\emptyset, \emptyset)  \quad
\stackrel{\small \textcircled{\raisebox{-0.9pt}{1}}}{\rightarrow} \;
 (\emptyset, \{(\textit{``user''}, 42) \mapsto \emptyset\}) \\
\stackrel{\small \textcircled{\raisebox{-0.9pt}{2}}}{\rightarrow} \;
& (\emptyset, \{(\textit{``user''}, 42) \mapsto \{\textit{login} \mapsto \textit{``hhiker''}\}\}) \\
\stackrel{\small \textcircled{\raisebox{-0.9pt}{3}}}{\rightarrow} \;
& (\emptyset, \{(\textit{``user''}, 42) \mapsto \{\textit{login} \mapsto \textit{``hhiker''}, 
                           \textit{pwd} \mapsto \textit{``galaxy''}\}\})  \\
\stackrel{\small \textcircled{\raisebox{-0.9pt}{4}}}{\rightarrow} \;
& (\{(\textit{``user''}, 42) \mapsto \{\textit{login} \mapsto \textit{``hhiker''}, 
                           \textit{pwd} \mapsto \textit{``galaxy''}\}\},\\
& 
\{(\textit{``user''}, 42) \mapsto \{\textit{login} \mapsto \textit{``hhiker''}, 
                           \textit{pwd} \mapsto \textit{``galaxy''}\}\}) 
\end{align*}

\vspace{-5mm}
$\punto$
\end{example}

\paragraph{Accessing entity values.} To access a particular value
of an entity, we introduce a dedicated operation.
We consider all variables as in Figure~\ref{fig:semantics},
with~$v$ being a property value. If such a value exists,
$v$ is either in $\textit{Dom}^+$
or a set of properties (from a nested entity):
$$
\llbracket \mbox{getProperty}(\kappa, n) \rrbracket (\textit{ds}, 
\textit{as}\cup \{\kappa \mapsto (\{n \mapsto v\} \cup \pi)\}) = v.
$$
If property~$n$ is not defined for the entity
with key~$\kappa$, calling~getProperty($\kappa$, $n$) yields~$\bot$.

\begin{example} \em  \label{ex:update_nested_entity}
We illustrate nesting and unnesting of entities
in close accordance with existing APIs (c.f.~\cite{google_datastore}).
Let $\kappa$ be the key of an entity with a nested entity as property~$n$.
To add a further property~$m$ with value~$w$ to the nested entity,
we unnest property~$n$ into a temporary entity.
Let  $\textit{tmp}$ be a new entity key.
After modification and re-nesting, we can persist the changes.
\begin{tabbing}
\quad \= get($\kappa$); \\
\> new($\textit{tmp}$, getProperty($\kappa$, $n$)); \quad \= {\em // unnesting}\\
\> setProperty($\textit{tmp}$, $m$, $w$);\\
\> setProperty($\kappa$, $n$, $\textit{tmp}$); \> {\em // nesting }\\
\> put($\kappa$)
\end{tabbing}

\vspace{-5mm}
$\punto$
\end{example}

\subsection{Queries}
\label{sec:queries_over_properties}

Given an entity key~$\kappa$, we define the function $\textit{kind}(\kappa)$
such that it returns the kind of this entity.
Then rule~\ref{sem:query_kind} retrieves all entities from the store
that are of the specified kind~$c$.

In addition to querying for a particular kind, we can also query with a
predicate~$\theta$,
as described by rule~\ref{sem:query_theta}.
We consider conjunctive queries, with equality as the only
comparison operator. This type of queries is typically supported
by all of today's NoSQL data stores. Various systems 
may even have more expressive query languages (e.g.\
with additional comparison operators and support for 
disjunctive queries, yet typically no join).

More precisely, $\theta$ is a conjunctive query over atoms of the form $n = v$
where $n$ is a property name and $v$ is a property
value from $\textit{Dom}$.
The predicate~$\theta$ is evaluated on one entity at-a-time.
We evaluate an atom $\kappa.n = v$ on a single entity:
\begin{align*}
 \llbracket n =  v \rrbracket ( \kappa \mapsto \pi) = &
\left\{ 
  \begin{array}{l l}
   \mbox{true} & (n \mapsto v) \in \pi\\
   \mbox{true} & (n \mapsto \vartheta) \in \pi, \vartheta \in \textit{Dom}^+, 
                  v \in \vartheta\\
    \mbox{false}  &  \text{otherwise}
  \end{array} \right.
\end{align*}
An atom involving~$\bot$ as an operand is always evaluated to false.
Queries over nested entities are not supported,
e.g.\ as in~\cite{google_datastore}.
The evaluation of conjunctions follows naturally:
$$
\llbracket \theta_1 \land \theta_2 \rrbracket (\kappa \mapsto \pi)
= \llbracket \theta_1 \rrbracket (\kappa \mapsto \pi) \land
\llbracket \theta_2 \rrbracket (\kappa \mapsto \pi)
$$

\subsection{Iteration Statements}
\label{sec:programmative_migration}

For batch updates on entities,
we define a for-loop.
Let~$x$ be a variable denoting an entity key,
and let~$\textit{op}$
denote an operation from our NoSQL database programming language.
$\theta$ is a conjunctive query with atomic equality conditions.
The operands in atoms are of the form~$x$, $x.n$, or~$v$,
where  $x$ is a variable denoting a key, $n$ is a property name,
and $v$ is a value from $\textit{Dom}$.

We consider  the execution of for-loops on a data store
state~$\textit{ds}$ and an application state~$\textit{as}$:
$$
\llbracket \mbox{\bf foreach } x  \mbox{ {\bf in} } 
\mbox{get($\theta$) }
\mbox{\bf do } \textit{op} \mbox{ {\bf od} }\rrbracket  (\textit{ds}, \textit{as}) \\
$$
Let $\textit{as}_\theta$ be the result of evaluating query~$\theta$,
i.e.\
$$
\llbracket \mbox{get($\theta$)}\rrbracket(\textit{ds}, \emptyset) = (\textit{ds}, \textit{as}_\theta)
$$
and let $K = \{\kappa \mid (\kappa \mapsto \pi) \in \textit{as}_\theta\}$
be the keys of all entities in the query result.
We can then evaluate the for-loop 
as follows.
\begin{tabbing}
\quad \= 
{\bf while} ($K \neq \emptyset)$ {\bf do} \\
\> \quad \= there exists some key $\kappa$ in $K$; \\
\> \>       $K := K \setminus \{\kappa \}$; \\[1mm]
\> \> evaluate operation $\textit{op}$ for the binding of~$x$ to key $\kappa$: \\
\> \> $(\textit{ds}, \textit{as}) := \llbracket \; \textit{op}[x/\kappa]\; \rrbracket
      (\textit{ds}, \textit{as})$;\\
\> {\bf od}
\end{tabbing}
Above,  $\textit{op}[x/\kappa]$ is obtained from operation~$\textit{op}$
by first substituting each occurrence of~$x$ in~$\textit{op}$
by~$\kappa$, and next replacing all operands $\kappa.n$
in query predicates by the value of ``getProperty($\kappa, n$)''.

\begin{example} \em \label{ex:safe_migrations}
We add a new property ``email'' to all user entities
in the data store, and initialize it with the empty string~$\epsilon$.
\begin{tabbing}
\;
   \= {\bf foreach} $x$ {\bf in} get($\textit{kind = \textit{``user''}}$) {\bf do} \\
      \> \quad \= setProperty($x, \textit{email}$, $\epsilon$);  \\
      \>       \>    put($x$)  \\
      \> {\bf od} 
\end{tabbing}
\noindent
Since denormalization is vital for performance in NoSQL data stores,
\nop{ 
we copy the property ``login'' from user entities
to new account entities.
We iterate over all user entities,
and define a local variable~$a$ denoting the key of a new
account entity. $\textit{id}: \textit{Keys} \rightarrow \textit{Id}$ is a function that extracts the identifier
from an entity key.
\begin{tabbing}
\;
   \= {\bf foreach} $u$ {\bf in} get($\textit{kind = \textit{``user''}}$) {\bf do} \\
      \> \quad \= Key $a$ = $(\textit{``account''}, \textit{id}(u))$; \quad {\em // A local variable}\\
      \>       \> new($a$); \\
      \>       \> setProperty($a$, $\textit{login}$, 
                                getProperty($u$, $\textit{login}$)); \\
      \>       \> put($a$)  \\
      \> {\bf od}
\end{tabbing}
}
we show how to copy  the property ``url'' from each user
entity to all blogposts written by that user.
\begin{tabbing}
\; \= 
{\bf foreach} $u$ {\bf in} get($\textit{kind}=\textit{``user''}$) {\bf do} \\
\>\quad\=  {\bf foreach} $b$ {\bf in} get($\textit{kind}=\textit{``blogpost''} \land 
                                      \textit{author} = u.\textit{login}$) {\bf do} \\
  \>  \> \quad \= setProperty($b$, $\textit{url}$, 
                    getProperty($u$, $\textit{url}$)); \\
\>\>\>    put($b$) \\
\>\> \bf od \\
\> \bf od
\end{tabbing}

\vspace{-5mm}
$\punto$
\end{example}

\section{Safe and Eager Migration }
\label{sec:encoding_evolution}

Now that we have a generic NoSQL database programming language,
we can implement the declarative schema evolution operations
from Section~\ref{sec:evolution}.
We believe the declarative operations cover common schema evolution tasks.
For more complex migration scenarios, we can always 
resort to a programmatic solution. This matches the situation
with relational databases, where an ``ALTER TABLE'' statement covers
the typical schema alterations, but where more complex 
transformations require an ETL-process to be set up, or a custom
migration script to be coded.

Figure~\ref{fig:implementing_operations_adr} shows the implementation
for the operations add, delete, and rename.
A for-loop fetches all matching entities from the data store,
modifies them, and
updates their version property (as introduced
in Section~\ref{sec:evolution}).
The updated entities are then persisted.

Figure~\ref{fig:implementing_operations_cm} shows the implementation
for copy and move. Again, entities
are fetched from the NoSQL data store one by one, updated,
and then persisted. This requires joins between
entities. Since joins are not supported in most NoSQL data stores,
they need to be encoded in the application logic.

This batch update corresponds to 
the recommendation of NoSQL data store providers
on how to handle schema evolution
 (e.g.~\cite{datastore_migration}).

Note that the create-or-replace semantics inherent in our NoSQL database
programming language make for a {\em well-defined}\/ behavior of operations.
For instance, renaming the property ``text'' in blogposts to ``content''
(c.f.\ Example~\ref{ex:rename}) effectively overwrites any existing property 
named content.

Moreover, the version property added to all entities makes the migration {\em robust}\/
in case of interruptions. NoSQL data stores commonly offer very limited transaction support.
For instance, Google Datastore only allows transactions to span up to five entities
in so-called {\em cross-group transactions}\/ (or alternatively, provides the concept
of entity groups not supported in our NoSQL database programming language)~\cite{google_datastore}.
So a large-scale migration cannot be performed as an atomic action. By restricting migrations to all entities of a 
particular version (using the where-clause), we may correctly recover from interrupts,
even for move and copy operations.

Interestingly, 
not all migrations that can be specified are desirable. For instance,
assuming a 1:N relationship between users and the blogposts they have written,
the result of the migration
$$
\mbox{{\bf copy} user.url {\bf to} blogpost {\bf where} user.login = blogpost.author}
$$
does not depend on the order
in which blogpost entities are updated. However, if there is an N:M relationship
between users and blogposts, e.g.\ since we specify
the copy operation as cross product between all users and all blogposts,
$$
\mbox{{\bf copy} user.url {\bf to} blogpost}
$$
then the execution order influences the migration result.
Naturally, we want to be able to know whether a migration is safe
before we execute it.
Concretely, we say a migration is {\em safe}\/ if it
does not produce more than one entity with the same key.%

\begin{figure}[t]
\paragraph{Legend:} Let $c$ be a kind, let $n$ be a property name,
and let $v$ be a property value from $\textit{Dom}$.
$\theta$ is  a conjunctive query over properties.\\[-2mm]

\underline{\bf add $c.n=v$ where $\theta$}
\begin{tabbing}
\quad \= {\bf foreach} $e$ {\bf in} get($\textit{kind}=c \land \theta$) {\bf do} \\
\> \quad \= setProperty($e$, $n$, $v$); \\ 
\> \>      setProperty($e$, $\textit{version}$, getProperty($e$, $\textit{version}$) $+ 1$); \\
\>\>       put($e$) \\
\> \bf od
\end{tabbing}

\underline{\bf delete $c.n$ where $\theta$}
\begin{tabbing}
\quad \= {\bf foreach} $e$ {\bf in} get($\textit{kind}=c \land \theta$) {\bf do} \\
\> \quad \= removeProperty($e$, $n$); \\ 
\> \>      setProperty($e$, $\textit{version}$, getProperty($e$, $\textit{version}$) $+ 1$); \\
\>\>       put($e$) \\
\> \bf od
\end{tabbing}

\underline{\bf rename $c.n$ to $m$ where $\theta$}
\begin{tabbing}
\quad \= {\bf foreach} $e$ {\bf in} get($\textit{kind}=c \land \theta$) {\bf do} \\
\> \quad \= setProperty($e$, $m$, getProperty($e$, $n$)); \\ 
\>  \> removeProperty($e$, $n$); \\ 
\> \>      setProperty($e$, $\textit{version}$, getProperty($e$, $\textit{version}$) $+ 1$); \\
\>\>       put($e$) \\
\> \bf od
\end{tabbing}
\caption{Implementing add, delete, and rename.}
\label{fig:implementing_operations_adr}
\end{figure}

\begin{figure}[t]

\paragraph{Legend:} Let $c_1, c_2$ be kinds and let $n$ be a property name.
Conditions~$\theta_1$ and~$\theta_2$ are conjunctive queries.
$\theta_1$ has atoms of the form
$c_1.m = v$, where $m$ is a property name and $v$ is a value from $\textit{Dom}$.
$\theta_2$ has atoms 
of the form $c_2.m = v$ or $c_1.a = c_2.b$, where  $a, b$, and 
$m$ are property names.
$v$~is a value from $\textit{Dom}$.\\[-2mm]

\underline{\bf move $c_1.n$ to $c_2$ where $\theta_1 \land \theta_2$}
\begin{tabbing}
\quad \= {\bf foreach} $e$ {\bf in} get($\textit{kind}=c_1 \land \theta_1$) {\bf do} \\
\quad \> \quad \= {\bf foreach} $f$ {\bf in} get($\textit{kind}=c_2 \land \theta_2$) {\bf do} \\
\> \> \quad \= setProperty($f$, $n$, getProperty($e$, $n$)); \\ 
\> \> \>     setProperty($f$, $\textit{version}$, getProperty($f$, $\textit{version}$) $+ 1$); \\
\>\> \>        put($f$) \\
\>\> \bf od; \\
\> \>     setProperty($e$, $\textit{version}$, getProperty($e$, $\textit{version}$) $+ 1$); \\
\> \>     removeProperty($e$, $n$); \\
\> \>        put($e$) \\
\> \bf od
\end{tabbing}

\underline{\bf copy $c_1.n$ to $c_2$ where $\theta_1 \land \theta_1$}
\begin{tabbing}
\quad \= {\bf foreach} $e$ {\bf in} get($\textit{kind}=c_1 \land \theta_1$) {\bf do} \\
\quad \> \quad \= {\bf foreach} $f$ {\bf in} get($\textit{kind}=c_2 \land \theta_2$) {\bf do} \\
\> \> \quad \= setProperty($f$, $n$, getProperty($e$, $n$)); \\ 
\> \> \>     setProperty($f$, $\textit{version}$, getProperty($f$, $\textit{version}$) $+ 1$); \\
\>\> \>        put($f$) \\
\>\> \bf od \\
\> \bf od
\end{tabbing}
\caption{Implementing copy and move.}

\label{fig:implementing_operations_cm}
\end{figure}

The following propositions follow from the implementations of 
schema evolution operators in
Figures~\ref{fig:implementing_operations_adr} 
and~\ref{fig:implementing_operations_cm}.

\begin{proposition}
An add, delete, or rename operation is safe.
\end{proposition}

\begin{proposition}
For a move or copy operation, 
and a data store state~$\textit{ds}$,
the safety of executing the operation on~$\textit{ds}$
can be decided in $O(|\textit{ds}|^2)$.
\end{proposition}

Deciding whether a copy or move operation is safe can be done in 
a simulation run of the evolution operator. 
If an entity has already been updated
in such a  ``dry-run'' and is to be overwritten with different property values,
then the migration is not safe.

In relational data exchange, the existence of solutions 
for relational mappings under constraints is a highly related problem.
There, it can be shown that while the existence of solutions
is an undecidable problem per-se, for certain restrictions,
the problem is PTIME-decidable~(c.f.\ 
Corollary~2.15 in~\cite{synthesis}). Moreover, 
the vehicle for checking for solutions is the chase algorithm,
which fails when equality-generating dependencies in the target
schema are violated. This is essentially the same idea
as our dry-run producing entities with the same key, but conflicting values.
Since our schema evolution operations copy and move
require two nested for-loops, we can check for safety in  quadratic time.
(Keeping track of which entities have already been updated
can be done efficiently, e.g.\ by maintaining
a bit vector in the size of $\textit{ds}$.)


\section{An Outlook on Lazy Migration}
\label{sec:lazy}
 
Our NoSQL database programming language can also express 
operations for lazy migration. 
To illustrate this on an intuitive level, we encode
some features of the Objectify object mapper~\cite{objectify_migration}.

We will make use of some self-explanatory additional language constructs,
such as if-statements and local variables. Additionally,
we assume an operation ``hasProperty($\kappa$, $n$)''
that tests whether the entity with key~$\kappa$
in the application state has a property by name~$n$.

\begin{example} \em
The following example is adapted from the
Objectify documentation.
It illustrates how properties are renamed
when an entity is loaded from the data store
and translated into a Java object.

The Java class Person is mapped to an entity.
The annotation~\verb!@Id! marks the identifier for this entity,
the entity kind is derived from the class name.
The earlier version of this entity has a property ``name'', which is 
now renamed to ``fullName''.
Legacy entities do not yet have the property ``fullName''.
When they are loaded, the object mapper assigns the value 
of property ``name'' to the class attribute ``fullName''.
The next time that the entity is persisted,
its new version will be stored.

{\small 
\begin{verbatim}
   public class Person {
     @Id Long id;
     @AlsoLoad("name") String fullName;
   }
\end{verbatim}
}

\noindent
In our NoSQL database programming language,
we implement the annotation \verb!@AlsoLoad! as follows.
\begin{tabbing}
\quad \= 
   Key $p$ := $(\textit{``Person''}, \textit{id})$; \\
\> {\bf if } hasProperty($p$, $\textit{name}$)  {\bf do } \\
\> \quad \= setProperty($p$, $\textit{fullName}$, getProperty($p$, $\textit{name}$)); \\
\> \>       removeProperty($p$, $\textit{name}$)\\
\> \bf od
\end{tabbing}

\vspace{-5mm}
$\punto$
\end{example}

\begin{example} \em
The following example is adapted  from~\cite{objectify_migration}.
The annotation \verb!@OnLoad!  specifies the migration 
for an entity when it is loaded.
If the entity has properties street and city,
these properties are moved to a new entity storing the address.
These properties are then discarded from the person entity
when it is persisted (specified 
by the annotation \verb!@IgnoreSave!).
Saving an entity is done
by calling the Objectify function \verb!ofy().save()!.

{\small 
\begin{verbatim}
   public class Person {
     @Id Long id;
     @IgnoreSave String street;
     @IgnoreSave String city;
  
     @OnLoad void onLoad() {
       if (this.street != null && this.city != null) {
         Entity a = new Entity("address");
         a.setProperty("person", this.id);
         a.setProperty("street", this.street);
         a.setProperty("city", this.city);
         ofy().save().entity(a);
        }  
     }  
   }
\end{verbatim}
}

\noindent
We implement the method with annotation \verb!@OnLoad! as follows.
\begin{tabbing}
\quad \= 
   Key $p$ := $(\textit{``Person''}, \textit{id})$; \\
\> {\bf if } ( hasProperty($p$, $\textit{street}$) $\land$ hasProperty($p$, $\textit{city}$) ) {\bf do } \\
\> \quad \= Key $a$ = $(\textit{``Address''}, \textit{id})$; \\
\> \>       new($a$); \\
\> \>       setProperty($a$, $\textit{person}$, $\textit{id}$); \\
\> \>       setProperty($a$, $\textit{street}$, getProperty($p$, $\textit{street}$)); \\
\> \>       setProperty($a$, $\textit{city}$, getProperty($p$, $\textit{city}$)); \\
\> \>       put($a$); \\
\> \>       removeProperty($p$, $\textit{street}$);\\
\> \>       removeProperty($p$, $\textit{city}$);\\
\> \bf od
\end{tabbing}

\vspace{-5mm}
$\punto$
\end{example}
%

It remains  future work to explore
lazy migrations in greater detail, and develop
mechanisms to statically check them prior to execution:
The perils of using such powerful features 
in an uncontrolled manner,
on production data, are evident.
Lazy migration is particularly difficult to test prior to launch,
since we cannot foretell which entities will be touched
at runtime. After all, users may return after years
and re-activate their accounts, upon which the object mapper
tries to evolve ancient data. 

It is easy to imagine scenarios where lazy migration fails,
due to artifacts in the entity structure that developers
are no longer aware of.
In particular, we would like to be able to determine whether
an annotation for lazy migration is safe.
At the very least, we would like to check
whether a lazy migration is {\em idempotent},
so that when transactions involving evolutions fail,
there is no harm done in re-applying the migration.


\section{Related Work}
\label{sec:related}

We define a NoSQL database programming language
as an abstract interface for programming against NoSQL data stores.
In recent work, \cite{simeon}~present a calculus for NoSQL systems
together with its formal semantics.
They introduce  a Turing-complete language and its type system, 
while we present 
a much more restricted language with a 
 focus on updates and schema evolution.

For relational databases, 
the importance of designing database programming languages
for strong programmability, concerning both performance and usability,
has been emphasized in~\cite{nextgen}.
The language presented  there 
can express database operators, query plans, and also capture
operations in the application logic.
However, the work there is targeted at query execution in
 relational databases, while we cover 
aspects of data definition and data manipulation in NoSQL data stores.
Moreover, we treat the data store itself as a black box,
assuming that developers use a cloud-based database-as-a-service
offering that they cannot manipulate.

All successful applications age with time~\cite{Parnas94}, 
and eventually require maintenance or evolution.
Typically, there are  two alternatives to handling this problem
on the level of schema: 
Schema versioning and schema evolution. 
Relational databases have an established language for schema 
evolution (``ALTER TABLE''). 
This schema definition language  is part of the SQL standard, 
and is implemented by 
all available relational databases systems. 

For evolving XML-based applications, 
research prototypes have been built that concentrate on the
 co-evolution of XML schemas and the associated XML 
documents~\cite{Guerrini2007}. 
The authors of~\cite{Necasky2012} 
have developed a model driven approach for XML schema design,
 and support co-evolution between different abstraction levels.
A dedicated language for XML evolution is introduced in~\cite{codex_language}
that formalizes XML  schema change operations and describes the corresponding  
updates of associated  XML documents.

JSONiq is a quite new query language for JSON documents, the first version was published in April 2013 \cite{jsoniq}.
Future versions of JSONiq will contain an update facility and will offer operations to add, delete, insert, rename, and replace properties and values. 
Our schema evolution language can be translated into corresponding update expressions. 
If JSONiq establishes itself as a standard for querying and updating NoSQL datastores, we can also base our schema evolution method on this language. 

The question  whether an evolution is safe
corresponds to the existence of (universal) solutions in data exchange.
In particular, established practices from 
XML data exchange, using regular tree grammars to specify the source and the
target schema~\cite{synthesis}, are highly relevant to our work.
The use of object mappers 
translating objects from the application space into
persisted entities can be seen as a form of schema specification.
This raises an interesting question: Provided that all entities
conform to the class hierarchy specified by an object mapper,
if we evolve entities, will they still work with our object mapper?
This boils down to checking for absolute consistency
in XML data exchange~\cite{synthesis}, and 
is a current topic in database theory (e.g.~\cite{abcons2}).
It is therefore  part of our plans to see how
we can leverage the latest research on
XML data exchange for evolving data in schema-less data stores.

There are various object-relational mapping~(ORM) 
frameworks fulfilling well established standards 
such as  the Java Persistence API~(JPA),
 and supporting almost all relational database systems.
Some ORM mappers are even supported by  NoSQL data stores,
of course not implementing all features,
since joins or foreign-keys are not supported by the backend
(e.g.\ see the~JPA and~JDO implementations for 
Google Datastore~\cite{google_jpa, google_jdo}).

So far, there are only few dedicated mappers for persisting objects
in NoSQL data stores (sometimes called object-data-store mappers (ODM)). 
Most of today's  ODMs are proprietary,
 supporting a particular NoSQL data store  
(e.g.\ Morphia~\cite{morphia} for MongoDB, 
or Objectify\cite{objectify} for Google Datastore).
Few systems support more than one NoSQL data store (e.g.\ Hibernate OGM~\cite{hibernate-ogm}).

Today, these objects-to-NoSQL mapping tools have 
at best rudimentary support for schema evolution. 
To the best of our knowledge, Objectify and Morphia
go the furthest by allowing developers to specify lazy migration
in form of object annotations.
However, we could not yet find any 
solutions for systematically managing and expressing
 schema changes. At this point,
the ecosystem of  tools for maintaining
NoSQL databases
is still within its infancy.



\section{Summary and Future Work}
This work investigates the maintainability
of feature-rich, interactive web applications,
from the view-point of schema evolution.
In particular, we target applications that are backed
by schema-less {\em document stores}\/ 
or {\em extensible record stores}\/.
This is an increasingly popular software stack,
now that 
database-as-a-service offerings are readily available:
The programming APIs are easy to use, there
is near to no setup time required, and pricing is reasonable.
Another sweet spot of these systems is
that the data's schema does not have to be specified
in advance. Developers may freely adapt the data's structure
as the application evolves.
Despite utter freedom,
the data  nevertheless displays an {\em implicit}\/ structure:
The application class hierarchy 
is typically reflected in the persisted data,
since object mappers perform  the mundane task 
of marshalling data between the application and the data store.

As an application evolves, so does its schema. 
Yet schema-free NoSQL data stores do not yet come with convenient
schema management tools. As of today, 
virtually all data migration tasks 
require custom programming (with the exception
of very basic data inspection tools for manipulating {\em single}\/ entities).
It is up to the developers to code the  migration of
their production data ``on foot'', getting the data  ready for the next
software release.
Worse yet, with weekly releases,
the schema evolves just as frequently.

In this paper, we lay the foundation for systematically
managing schema evolution in this setting.
We define a declarative {\em NoSQL schema evolution language}\/,
to be used in a NoSQL data store administration console.
Using our evolution language, developers can specify
common operations, such as adding, deleting, or renaming properties
in batch. Moreover,  properties can be  moved
or copied, since data duplication and denormalization
are fundamental in NoSQL data stores.
We emphasize that we do not mean to enforce
a relational schema onto NoSQL data stores.
Rather, we want to ease the pain of schema evolution
for application developers.

We regard it as one of our key contributions that 
our operations can be implemented 
for a large class of NoSQL data stores. We show this by 
an implementation in a generic {\em NoSQL database programming language}.
We also discuss which operations can be applied safely, 
since non-deterministic migrations are unacceptable.

\paragraph{Future work.}
Our NoSQL schema evolution language specifies operations
that are executed {\em eagerly}\/, on all qualifying entities.
An alternative approach is to migrate entities {\em lazily}\/,
the next time  they are fetched into the application space.
Some object mappers already provide such functionality.
We believe that lazy evolution is still little understood,
and at the same time poses great risks
when applied erroneously. We will investigate how our NoSQL schema
evolution language may be implemented both safely {\em and}\/ lazily.
Ideally, a dedicated schema evolution management tool would 
allow developers to migrate data eagerly
for leaps in schema evolution, and to patch things up lazily
for minor changes.







\end{document}